# Detection and correction of the misplacement error in THz Spectroscopy by application of singly subtractive Kramers-Kronig relations

Valerio Lucarini[1,2], Yusuke Ino[3], Kai-Erik Peiponen[1] and Makoto Kuwata-Gonokami[3]

1) Department of Physics, University of Joensuu, P.O. Box 111, 80101, Joensuu, Finland

2) Department of Mathematics and Computer Science, University of Camerino, 62032 Camerino (MC), Italy

3) Department of Applied Physics, the University of Tokyo and Solution Oriented Research for Science and Technology (SORST), JST, 7-3-1 Hongo, Bunkyo-ku Tokyo, 113-8656, Japan

**ABSTRACT**

In THz reflection spectroscopy the complex permittivity of an opaque medium is determined on the basis of the amplitude and of the phase of the reflected wave. There is usually a problem of phase error due to misplacement of the reference sample. Such experimental error brings inconsistency between phase and amplitude invoked by the causality principle. We propose a rigorous method to solve this relevant experimental problem by using an optimization method based upon singly subtractive Kramers-Kronig relations. The applicability of the method is demonstrated for measured data on an *n*-type undoped (100) InAs wafer in the spectral range from 0.5 up to 2.5 THz.

# 1. INTRODUCTION

Recent advances in the femto-second laser technology have enabled us to investigate the dielectric functions of materials in the far infrared region thorough the time-domain terahertz (THz) spectroscopy [1]. In time-domain THz spectroscopy, one can measure the time-domain waveform of the THz pulse. By applying the Fourier transform to this temporal waveform, one can obtain the amplitude and the phase of the THz wave in frequency space. Thus, in the THz spectral region one can resolve the complex optical quantity of a medium both in transmission and reflection modes. Reflection measurement is especially useful when considering a medium that is opaque in the THz frequency range. Due to the experimental restrictions in the reflection measurement, one has to put a reference sample in the same position as the sample of interest, typically with a spatial precision of a few microns. This is a quite tedious and critical task, which means that the measured phase can be easily distorted by non-negligible systematic phase error. Various experimental techniques have been suggested to remove the misplacement error [2-4] but the problem has not been solved clearly by such experimental procedures. Recently, a numerical method based on maximum entropy model (MEM) has been successfully applied for the reduction of this misplacement error [5, 6]. The main conceptual problem in this method is that MEM is not relying on any physical principle, but rather on information mathematics. The principle of causality rule is the ground for the existence of Kramers-Kronig (K-K) relations in linear optics [5, 7, 8]. The extension to nonlinear optics has been recently developed [9-11] and experimentally tested [12]. The traditional use of K-K relations is based either on extraction of the real part from the measured imaginary part of complex optical quantity or *vice versa*. In linear optics it means usually that wavelength-dependent refractive index is obtained from transmission spectrum, or the phase of the reflectivity is obtained from reflection spectrum by appropriate K-K analysis. The characteristic integral structure of K–K relations requires the knowledge of the spectrum at a semi-infinite angular frequency range. Unfortunately, in practical spectroscopy only finite spectral range can be measured. Although various interpolation methods have been proposed, there is still a lack of one that can be applied with high precision to arbitrary spectra. Indeed, it has been shown that commercial software packages of spectrophotometers that make use of K-K relations may give qualitatively different results even though using identical input data [13], specifically in cases that the selected frequency range contains only the either side of the resonance structure. In particular, the problem of assessment of the real refractive index with the aid of band-limited extinction coefficient has been a notable issue in this field [14, 15]. In the context of linear and nonlinear optics, singly subtractive Kramers-Kronig relations (SSKK) and multiply subtractive Kramers-Kronig (MSKK) have been proposed in order to relax the limitations due to the finiteness of the data range, because a much better convergence of the integrals in the dispersion relations is realized [10, 11, 16-19]. Rigorous mathematical derivation of SSKK and MSKK can be found in [11, 16, 18, 19]. For a complete overview of the theory and applications of the conventional and subtracted K-K relations in optical material research, we refer to the recently published book by Lucarini et al. [19].

In this paper we show that SSKK relations provide a very useful tool to remove the misplacement error and thus make the measurements and the interpretation of THz signal more easy and reliable. The procedure is conceptually based on the causality principle only, so that no *ad hoc* assumptions or additional information about the mechanism of the dielectric response is needed.

Our paper is organized as follows. In section 2 we outline how singly subtractive Kramers-Kronig relations can be effectively used in the context of THz spectroscopy. In section 3 we exemplify our approach by presenting a relevant application on measured data on an *n*-type undoped (100) InAs wafer [20] in the spectral range from 0.5 up to 2.5 THz. In section 4 we present our conclusions.

## 2. TERAHERZ SPECTROSCOPY AND SUBTRACTED KRAMERS-KRONIG RELATIONS

The optical problem related to the misplacement of the sample is the error in the optical path length $\delta L$, which induces the phase error $\omega \delta L/c$ of the reflected field, where $c$ is the light velocity in vacuum and $\omega$ is the frequency of the radiation. This means that the true phase of the reflected THz field can be expressed by $\theta(\omega)_{true} = \theta(\omega)_{measured} + \alpha\omega$, where $\alpha = \delta L/c$. So it is important to find a general procedure to find the value of $\alpha$. We suggest the adoption of a procedure based on the self-consistency of the K-K relations analysis. Most typically, K-K analysis is used only *one-way*, i.e. the unknown part of optical function is derived from the measured or modelled part (most typically the imaginary part) of the same optical function via the appropriate dispersion relation.

### A. Singly Subtractive Kramers-Kronig Relations

The power of SSKK and MSKK over conventional K-K relations is the better convergence of the integrals [10, 11, 16-19], which means that the requirement of data on semi-infinite frequency range is relaxed. We wish to emphasize that the present method is not restricted to any particular medium or theoretical model for the spectra but it is totally general. In the case of SSKK analysis the dispersion integrals are as follows:

$$\ln|r(\omega)| - \ln|r(\omega_1)| = \frac{2(\omega^2 - \omega_1^2)}{\pi} P \int_0^\infty \frac{\omega' \theta(\omega')}{(\omega'^2 - \omega^2)(\omega'^2 - \omega_1^2)} d\omega' \qquad (1)$$

$$\omega^{-1}\theta(\omega) - \omega_2^{-1}\theta(\omega_2) = -\frac{2(\omega^2 - \omega_2^2)}{\pi} P \int_0^\infty \frac{\ln|r(\omega)|}{(\omega'^2 - \omega^2)(\omega'^2 - \omega_2^2)} d\omega', \qquad (2)$$

where $r(\omega)$ is the complex reflection coefficient, $\omega_1$ and $\omega_2$ are anchor points and P denotes the Cauchy principal value. In the presence of mismatching error the phase cannot be perfectly retrieved from the reflectance and *vice versa* by using SSKK relations (1) & (2). That is to say since we have measured both the amplitude and the phase of the reflectivity, we can use simultaneously both relations (1) & (2) in order to check if the data is in accordance with these relations. Departures between measured (i.e. including phase error) and calculated data indicate that there should be an error in the measured data. The SSKK and more generally MSKK require always the consistency between measured and inverted data.

By applying the SSKK relations (1) and (2) to the measured $\ln(|r(\omega)|)$ and $\theta(\omega)$ (including phase error) we cannot obtain any positive match. We emphasize that usually the reflectivity is subject to minor change whereas the misplacement error induces a large error in the phase of the reflected electric field thus the measured phase departs to great extent from the correct phase.

## B. The Optimization Procedure

We find the correct phase by an optimization process where both relations (1) & (2) are exploited simultaneously and we let the parameter $\alpha$ as well as the location of the anchor points vary. The idea is that when we capture the appropriate phase correction we have self-consistency between the data obtained by the SSKK analysis. The optimization procedure can be summarized as in the flow chart depicted in figure 1, which is fully described in this paragraph. Here we select the anchor points with the simplifying assumption $\omega_1 = \omega_2$. The first action is to initialize the optimisation procedure by guessing a correction to the experimental phase error obtained by adding a linear term in the form $\theta(\omega)_{corrected} = \theta(\omega)_{measured} + \alpha\omega$ (steps 1 to 4). Within the loop, the SSKK relations (1) and (2) are applied to improve the estimates of the true values of the functions $\ln(|r(\omega)|)$ and $\theta(\omega)$ (steps 5 to 9). The loop is repeated until the $L^2$ distance $|r(\omega)_{guess} - r(\omega)_{corrected}|$, which gives a measure of the degree of self-consistency of the SSKK data inversion procedure, has an incremental improvement smaller than ε with respect to the previous iteration (step 10). When the loop ends, we derive $r(\omega)_{selfcons} = \exp[\ln(|r(\omega)|)_{selfcons} + i\theta(\omega)_{selfcons}]$ (step 11).

Since the goodness of the match is evaluated by computing the $L^2$ norm of the complex function $|r(\omega)_{selfcons} - r(\omega)_{corrected}|$, by minimizing such norm with respect to $\alpha$, we thus obtain a value $\alpha_{optimal}$. Such value of $\alpha$ indicates the best correction to the measured phase, where best is in the sense of adherence to the dispersion relations SSKK (1) and (2) in terms of the $L^2$ norm. The choice of this norm is especially appropriate in mathematical terms when considering the outcomes of dispersion relations [8, 19]. Since the choice of the anchor point $\omega_1$ is not intrinsic in the theory, while being critical in practical terms, we expect that the estimate for the actual α is robust if $\alpha_{optimal}$ shows a weak dependence on $\omega_1$. Similar observation holds also for the optimization parameter ε (in this case chosen as ε=0.01).

# 3. APPLICATION ON *n*-TYPE UNDOPED (100) InAs WAFER

We show an example of the application of this procedure. Here we treat the case of THz time-domain reflection measurement on *n*-type undoped (100) InAs wafer in the spectral range from 0.5 up to 2.5 THz. Detailed description of the sample and experiments can be found in previous papers [6, 20]. Following the above described procedure, we obtain the $L^2$ norm-optimal value of α, indicated by $\alpha_{optimal}$, for each choice of $\omega_1$. In Figure 2 we observe that the value of $\alpha_{optimal}$ is well defined in the whole domain of $\omega_1$ and especially well for values $\omega_1 \in$ [1THz, 2THz], such that the anchor point is relatively far from the boundaries. By computing the statistics with respect to $\omega_1$, we obtain a mean value and a standard deviation of the distribution of $\alpha_{optimal}$, which are also depicted in Figure 2. The mean value of $\alpha_{optimal}$ thus constitutes our best estimate of the misplacement correction coefficient α.

As in the previous paragraph, we can estimate $r(\omega)_{corrected}$ and $r(\omega)_{selfcons}$ if we set the parameters $\alpha$ and $\omega_1$. It is important to estimate the influence of the statistical uncertainty of $\alpha$ on the estimated $r(\omega)_{corrected}$ and $r(\omega)_{selfcons}$. In Figure 3a) we show the self-consistent SSKK results obtained for the absolute value of the reflectance. In Figure 3b) we plot the mean value and the standard deviation of $|r(\omega)|_{selfcons}$ versus $\omega$. These statistics are also computed with respect to $\omega_1$, assuming that $\alpha_{optimal}$ obeys the statistics described in Figure 2. We see that the measured value of the reflectance fits well with the self-consistent estimates over the whole range of frequency $\omega$, except the very boundaries. This confirms that the measurement of the reflectance is *robust* in spite of the misplacement error so that the measured reflectance $|r(\omega)|_{measured}$ can be taken, within experimental errors, as the true reflectance of the sample.

We depict in Figure 4a) the results obtained for the value of $\theta(\omega)_{selfcons}$ and in Figure 4b) we depict its mean value and standard deviation with respect to $\omega_1$. In Figure 5a) we depict instead the best estimate obtained for the corrected phase $\theta(\omega)_{measured} + \alpha_{optimal}\omega$. In Figure 5b) we depict its mean and standard deviation, where again the statistics is computed with respect to $\omega_1$, assuming again that $\alpha_{optimal}$ obeys the statistics described in Figure 2. The correction with respect to the measured phase is large, which suggests the importance of the procedure of removing the misplacement error. Comparing Figures 4b) and 5b), we observe that the match between the corrected and the self-consistent estimate of the phase is excellent for $\omega \leq 2$THz, while a relatively large disagreement is found in the upper portion of the considered spectrum. Since we know that the misplacement causes an error in the evaluation of the phase such that $\theta(\omega)_{true} = \theta(\omega)_{measured} + \alpha\omega$, we indicate that the estimate of the corrected phase $\theta(\omega)_{measured} + \alpha_{optimal}\omega$ presented in Figure 4, rather than the self-consistent estimate phase $\theta(\omega)_{selfcons}$ presented in Figure 3 provides the best approximation of the true phase.

## 4. SUMMARY AND CONCLUSIONS

We have presented a general procedure for obtaining the phase error $\alpha_{optimal}\omega$ and the misplacement error $\delta L = c\alpha_{optimal}$ of a time-domain THz-reflectivity measurement. We have used a self-consistent procedure based on SSKK, and derived the optimal value of $\alpha$ by imposing the best match between self-consistency in terms of the dispersion relation, and corrected measured data. We have shown that the results obtained are robust since they are rather similar within a large domain of choice of the anchor point. Moreover, the consideration of various anchor points, which is enabled in time-domain THz spectroscopy, allows us to pose the retrieval problem in statistical terms and obtain uncertainty bounds on the estimated optimal parameters. We have provided the analysis of THz time-domain reflection measurements on *n*-type undoped (100) InAs wafer in the spectral range from 0.5 up to 2.5 THz as an example. The proposed method drastically simplifies the THz reflection spectroscopy. Our algorithm is applicable to the case that the sample is sealed with optically opaque materials, or has a relief structure of a few microns on its surface. Finally, it should be noted that the developed procedure can be extended to other frequency region including the optical frequency by the combination with recently developing ultrafast phase-sensitive optical technology [21]. One can apply the technique for the diagnostics of the dispersive optical components such as chirped Bragg mirrors.

**Acknowledgments**

The authors wish to thank two anonymous reviewers for useful comments. V. L. and K.-E. P. wish to thank the Academy of Finland for Grant no. 204109.Y.I and M.K.G. wish to thank R. Shimano for the fruitful discussions and also thank the support by JSPS KAKENHI(S).

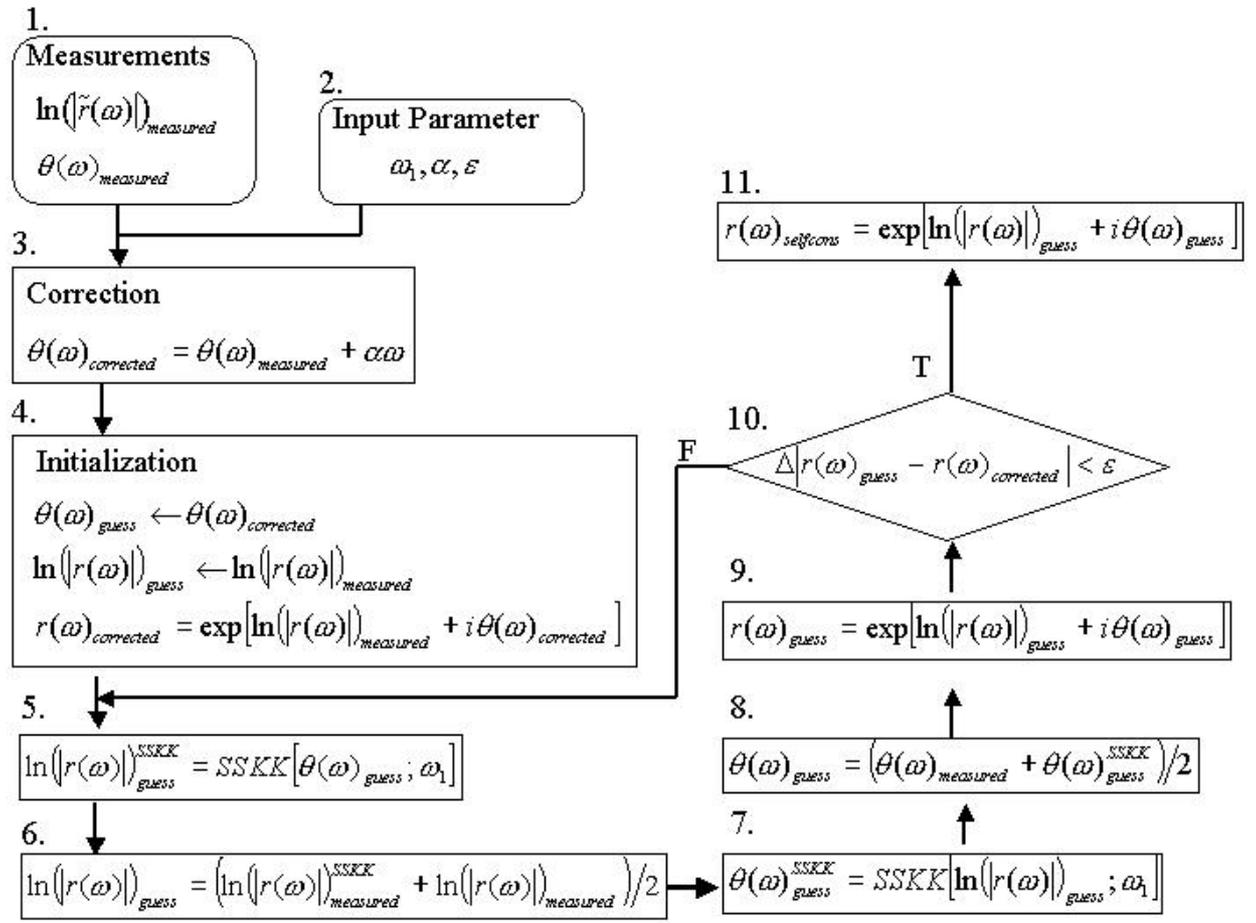

**Figure 1**

Flow chart of the optimization procedure. We note that the rhombic figure stands for a logic statement: Δ refers to relative improvement with respect to the previous iteration, T (true) indicates the flow in the case the statement is true, F (false) in the case it is false.

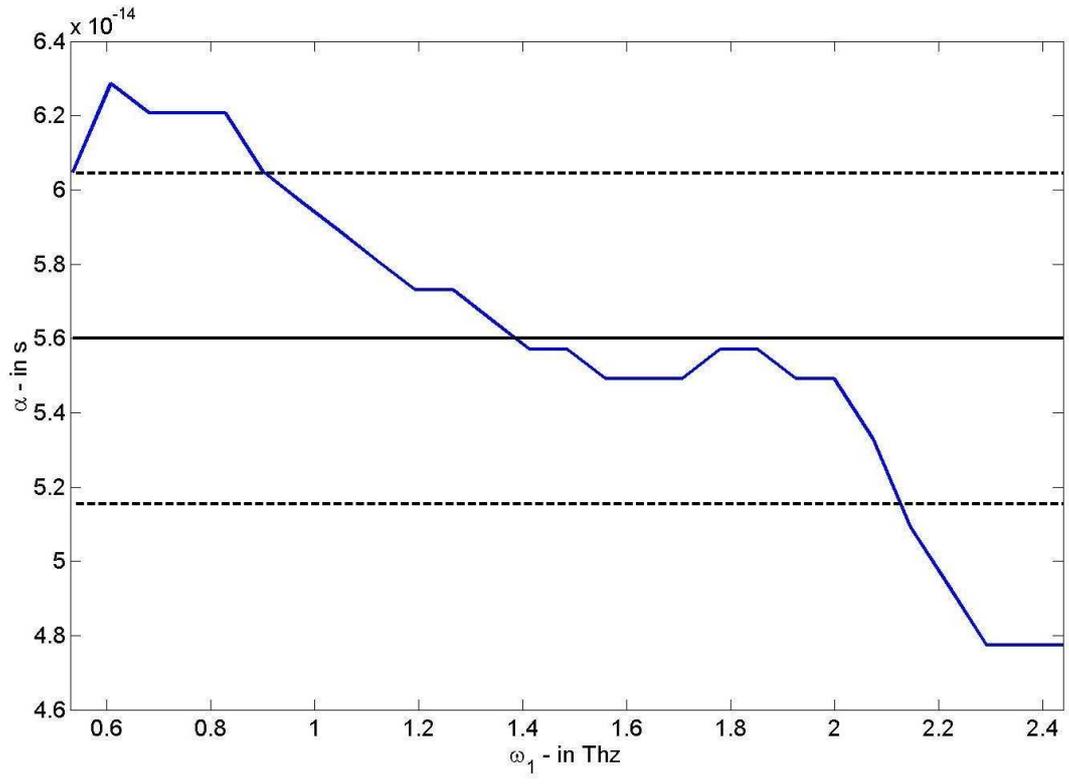

Figure 2: Value of $α_{optimal}$ (blue line) as a function of $ω_1$. The mean (black solid line) ± the standard deviation (black dashed lines) are also indicated.

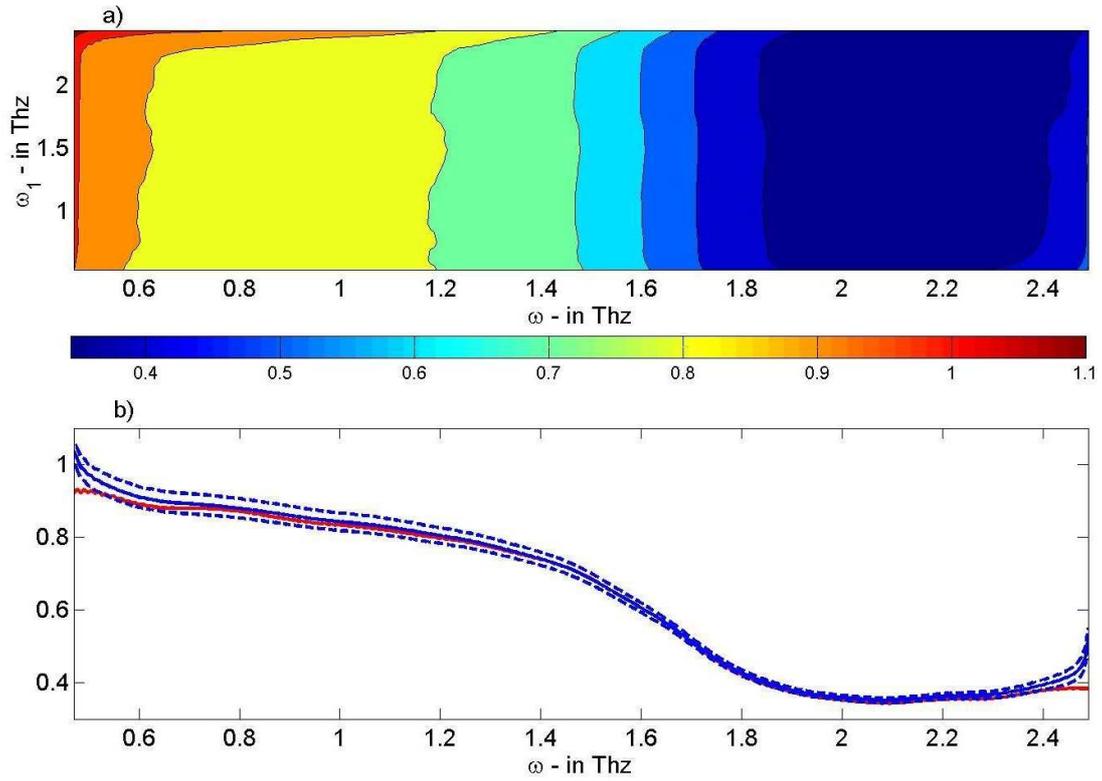

**Figure 3** Reconstruction of the value of the reflectance. a) SSKK self-consistent estimate of $|r(\omega)|_{selfcons}$ as a function of $\omega_1$. b) Measured $|r(\omega)|_{selfcons}$ (red solid line), mean value (blue solid line) ± standard deviation (blue dashed lines) of the best estimate of $|r(\omega)|_{selfcons}$ computed with respect to $\omega_1$.

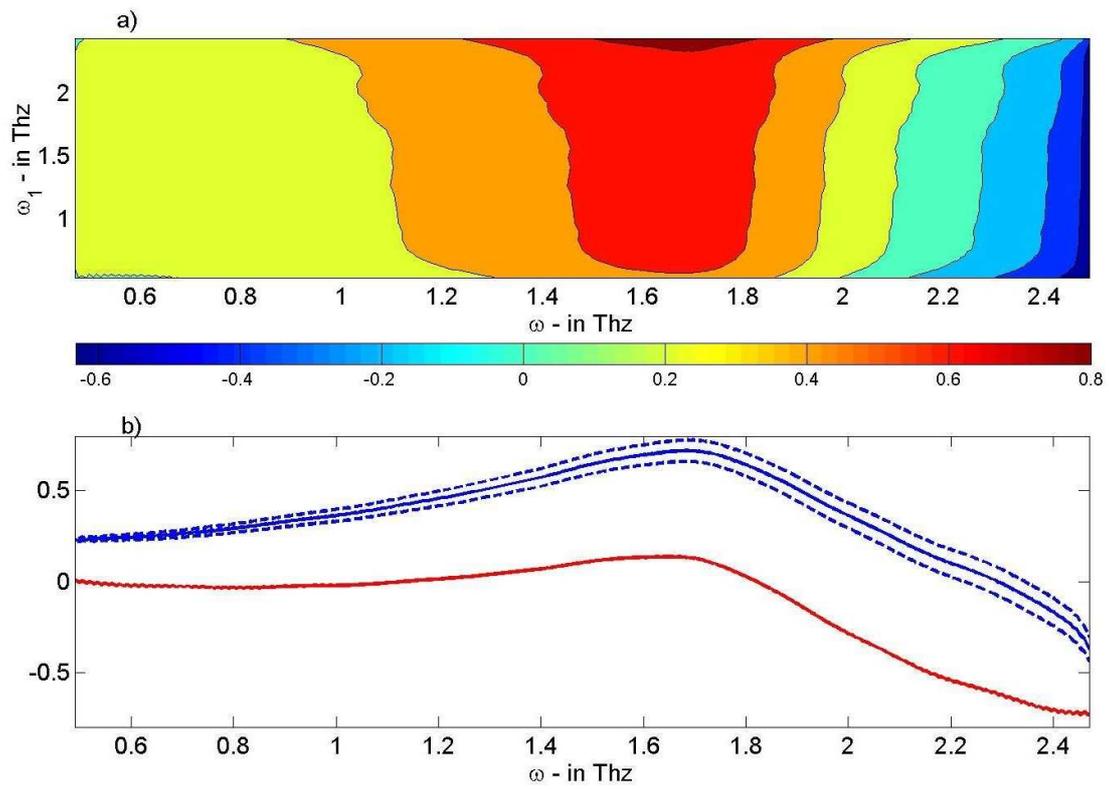

**Figure 4 Reconstruction of the value of the reflectance. a) SSKK self-consistent estimate of $\theta(\omega)_{selfcons}$ as a function of $\omega_1$. b) Measured $\theta(\omega)_{selfcons}$ (red solid line),, mean value (blue solid line) ± standard deviation (blue dashed lines) of the best estimate of $\theta(\omega)_{selfcons}$ computed with respect to $\omega_1$.**

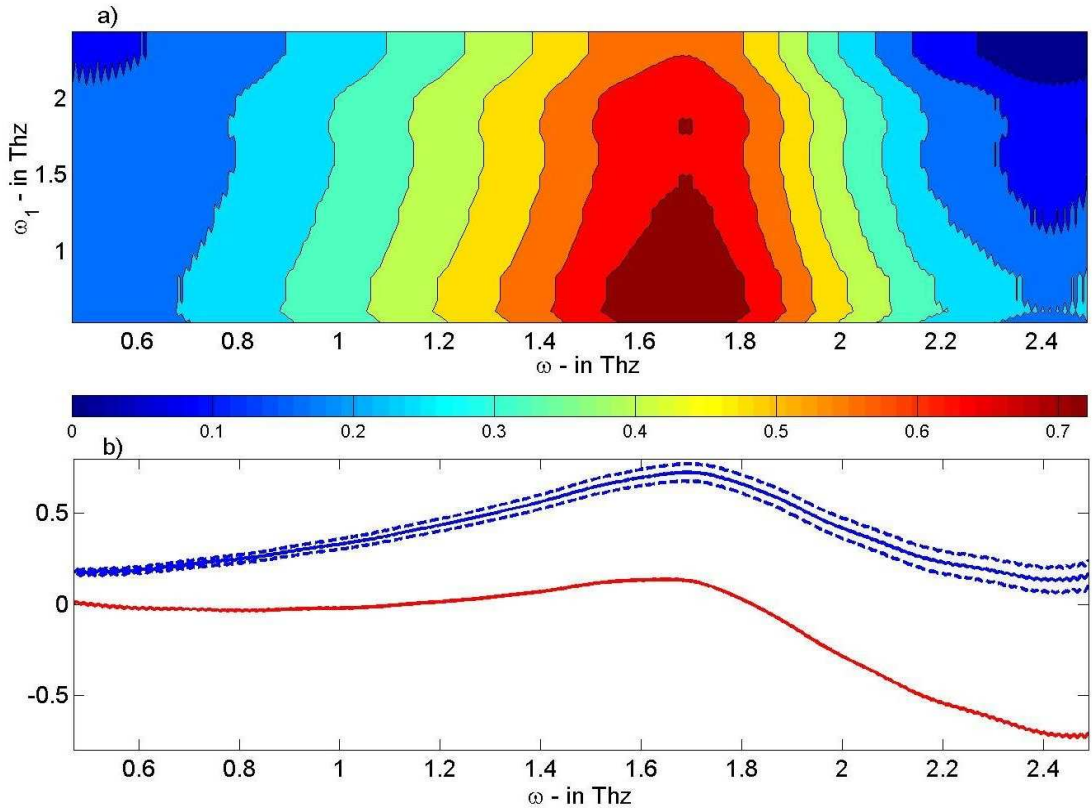

**Figure 5 Reconstruction of the phase of the reflectance. a) Best estimate of the corrected phase** $\theta(\omega)_{measured} + \alpha_{optimal}\omega$ **as a function of** $\omega_1$**. b) Measured phase (red solid line),, mean value (blue solid line) ± standard deviation (blue dashed lines) of the best estimate of the actual phase** $\theta(\omega)_{measured} + \alpha_{optimal}\omega$ **with respect to** $\omega_1$**.**